\documentclass[prl,
superscriptaddress,a4paper, twocolumn,
showpacs,amsmath,amssymb]{revtex4}
\usepackage{units}
\usepackage{graphicx}

\begin{document}

\title{
Transition from a Tomonaga-Luttinger liquid to a Fermi liquid in
potassium intercalated bundles of single wall carbon nanotubes}
\author{H. Rauf}
\author{T. Pichler}
\author{M. Knupfer}
\author{J. Fink}
\affiliation{Leibniz-Institut f\"ur Festk\"orper- und Werkstoffforschung,
  D-01069 Dresden, Germany}
\author{H. Kataura}
\affiliation{Nanotechnology Research Institute, National Institute
of Advanced Industrial Science and Technology (AIST) 1-1-1
Higashi, Tsukuba, Ibaraki 305-8562, JAPAN}

\bigskip

\begin{abstract}

We report on the first direct observation of a transition from a
Tomonaga-Luttinger liquid to a Fermi liquid behavior in potassium
intercalated mats of single wall carbon nanotubes (SWCNT). Using
high resolution photoemission spectroscopy an analysis of the
spectral shape near the Fermi level reveals a Tomonaga-Luttinger
liquid power law scaling in the density of states for the pristine
sample and for low dopant concentration. As soon as the doping is
high enough to fill bands of the semiconducting tubes a distinct
transition to a bundle of only metallic SWCNT with a scaling
behavior of a normal Fermi liquid occurs. This can be explained by
a strong screening of the Coulomb interaction between charge
carriers and/or an increased hopping matrix element between the
tubes.

\end{abstract}


\maketitle

The charge transport properties of carbon nanotubes have been
investigated intensively over the last years since they represent
an archetype of a one dimensional system \cite{egger,kane,egger2}.
For such metallic systems, conventional Fermi-liquid (FL) theory
fails since even the smallest interaction between the charge
carriers leads to very strong correlation effects. Correlation
effects are one of the central research areas in solid state
physics and therefore one-dimensional metals are a paradigm for
solids, where a breakdown of the FL theory due to many-body
problems is expected. Under certain conditions a one-dimensional
metal forms a Tomonaga-Luttinger liquid (TLL) which shows peculiar
behavior such as spin charge separation and interaction dependent
exponents in the density of states, correlation function and
momentum distribution of the electrons
\cite{egger,kane,ll_rev,ll_rev2}. Results from transport
measurements through junctions between metals and individual
metallic carbon nanotubes as well as between carbon nanotubes have
been extensively analyzed in the framework of a tunnelling into or
between TLL \cite{bock,dekker,schoenenberger}. Recently, the
electronic density of states (DOS) of the valence band electrons
of mats of single wall carbon nanotubes (SWCNT) was directly
monitored by angle integrated high resolution photoemission
experiments \cite{tll_nature}. The spectral function and the
temperature dependence of the intensity at the Fermi level
exhibited a power law dependence with exponents of 0.46 and 0.48
respectively which are identical within experimental error. This
value yields a TLL parameter $g=0.18$ in very good agreement with
theoretical predictions \cite{egger,kane} and consistent with
transport experiments through carbon nanotubes between normal
metals \cite{schoenenberger}. This photoemission study clearly
evidenced that metallic SWCNT within a bundle of SWCNT can be
described within TLL theory regarding their low energy properties,
without uncertainties regarding their contacting. However, there
is still an open question regarding the amount of metallic SWCNT
within a bundle of SWCNT. The interaction within a bundle could
yield to the opening of a small gap in all SWCNT \cite{lieber}.
Hence all SWCNT would be narrow gap semiconductors which could not
be described within TLL theory. On the other hand, tight binding
calculations of bundles of metallic SWCNT pointed out that only
for a (10,10) crystal a pronounced pseudo gap of about 0.1 eV is
observed, whereas in a disordered bundle of metallic SWCNT with
different chirality the interaction between neighboring SWCNT is
weak and has a negligible effect on the DOS in the vicinity of the
Fermi level \cite{mele}.
\par
In this Letter we first address the question if the power law
behavior observed in the recent photoemission study
\cite{tll_nature} is related to the existence of the above
mentioned pseudo gap or to a TLL behavior. By means of doping, the
Fermi level for the metallic tubes could be shifted away from the
pseudo gap region and still a power law behavior with the same
$\alpha$ is observed. Secondly, it is interesting to study the
case where the Fermi level is shifted into the states of the
semiconducting tubes of the bundle. A strong reduction of $\alpha$
is observed due to the filling of a band from the semiconducting
tubes with non-one-dimensional character. Finally, at the highest
doping levels and probably due to the filling of non-linear
dispersing bands of the metallic tubes, a clear Fermi edge,
typical of a normal FL behavior, is observed.
\par
One very efficient possibility to change the electronic properties
by doping with electrons or holes is intercalation which has been
studied extensively for compounds such as fullerenes (FIC)
\cite{1} and graphite (GIC) \cite{dress1}. For SWCNT intercalation
compounds in contrast to FIC and GIC no distinct intercalation
stages have been observed as yet. Alkali metal intercalation of
mats of bundled SWCNT takes place inside the channels of the
triangular bundle lattice \cite{ssc1,liukx} and leads to a shift
of the Fermi energy, a loss of the optical transitions \cite{kaz}
and an increase of  the conductivity by about a factor of thirty
\cite{cond,ssc1}. A complete charge transfer between the donors
and the SWCNT was observed up to saturation doping, which was
achieved at a carbon to alkali metal ratio of about seven
\cite{ssc1,liukx}. However, much less has been reported on direct
measurements of the low energy electronic properties as a function
of doping. First results using photoemission revealed a Fermi edge
at high doping \cite{pes_prb}. Here, we report the first detailed
study of the change in the low energy electronic properties in
mats of SWCNT as a function of potassium intercalation using high
resolution photoemission as a probe.
\par
Mats of purified SWCNT which consist of a mixture of roughly 2/3
of semiconducting and 1/3 metallic SWCNT with a narrow diameter
distribution which is peaked at 1.37 nm with a variance of about
0.05 nm \cite{tll_nature} were produced by subsequent dropping of
SWCNT suspended in acetone onto NaCl single crystals. The produced
SWCNT film of about 500 nm thickness was floated off in distilled
water and recaptured on sapphire plates. For the photoemission
experiments the sample was mounted onto a copper sample holder and
cleaned in a preparation chamber under ultra high vacuum (UHV)
conditions (base pressure 9x10$^{-11}$ mbar) by electron beam
heating to 800 K. Electrical contact of the SWCNT film was
established by contacting the surface to the sample holder via a
Ta foil. Then the sample was cooled down to T=35 K and transferred
under UHV conditions to the measuring chamber and analyzed
regarding the electronic properties using a hemispherical high
resolution Scienta SES 200 analyzer. For the angle integrated
valence band photoemission spectra using monochromatic HeI$\alpha$
(21.22 eV) excitation the energy resolution was set to 10 meV. The
core level photoemission measurements (XPS) were performed at 400
meV energy resolution using monochromatic AlK$\alpha$ excitation
(1486.6 eV). The Fermi energy and overall resolution was measured
on freshly cleaned Ta. The intercalation was performed  {\it in
situ} after heating the sample to 450 K using commercial SAES
potassium getter sources. After subsequent exposure to the dopant
vapor an additional equilibration for about 30 min at 450 K was
performed to increase the sample homogeneity.
\par
The sample stoichiometry and purity was checked by core level
photoemission spectroscopy. No contamination from oxygen or
catalyst particles could be detected. The binding energy of the
C1s line is shifted by about 0.8 eV to higher values for the
highest dopant concentration (here C/K = 15) This can be explained
by an upshift of the Fermi level into the conduction band in good
agreement with results from electron energy-loss and Raman
spectroscopy \cite{liukx,peaprl} and similar to the corresponding
GIC\cite{oel}. The doping level was determined by the ratio of the
C1s/K2p intensities taking into account the different
photo-ionization cross sections.

\begin{figure}[htb]
\includegraphics[width=0.68\linewidth]{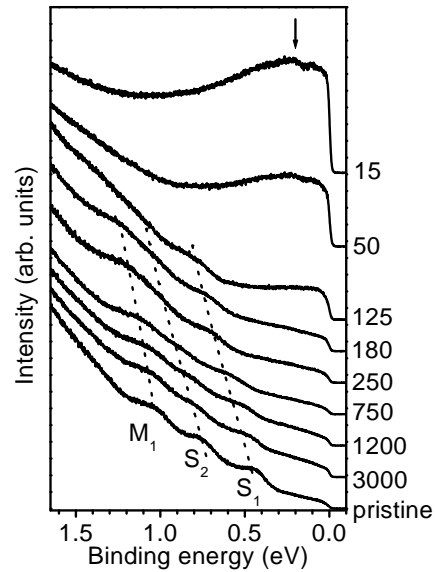}
\caption{Doping dependence of the valence band photoemission
spectra in the vicinity of the Fermi edge for high doping levels.
The numbers correspond to the C/K ratio derived from core level
photoemission. The dotted lines are guidelines for the evolution
of the S$_1$, S$_2$ and M$_1$ peaks with increasing doping. The
arrow highlights the satellite in the photoemission response at
high doping levels.}
\end{figure}

We now turn to the detailed analysis of the high resolution
valence band photoemission experiments at T=35K. The results are
depicted in Fig. \ 1. Compared to graphite, only close to the
Fermi level distinct differences are observed in the photoemission
response of the SWCNT mats \cite{tll_nature,pes_un}. For the
pristine SWCNT, the photoemission peaks corresponding to the first
and second van Hove singularity (vHs) of the semiconducting SWCNT
(S$_1$,S$_2$) and that of the first vHs of the metallic SWCNT
(M$_1$) are observed  - in very good agreement with previous
results \cite{tll_nature} - at binding energies of 0.44, 0.76 and
1.06 eV, respectively. In the simplest picture the valence band
vHs S$_1$ and the conduction band vHs S$_1^\star$ would be
symmetric around the Fermi level of the SWCNT bundle yielding a
S$_1^\star$ position of 0.44 eV above the Fermi level. However, it
is well known from scanning tunnelling spectroscopy \cite{sts} of
individual nanotubes that the vHs peaks of semiconducting SWCNT
can be shifted away up to 0.1 eV with respect to the Fermi level
due to charge carrier injection from the substrate. A similar
effect can be expected for our photoemission experiments due to a
redistribution of the charges within a bundle consisting of a
mixture of semiconducting and metallic SWCNT by contact
potentials. This allows us to safely estimate the position of the
conduction band S$_1^\star$ being at least 0.3 eV above the Fermi
level of the SWCNT bundle. With increasing doping, the peaks
corresponding to the SWCNT vHs (S$_1$, S$_2$, and M$_1$) shift to
higher binding energy due to a filling of the conduction band of
the SWCNT with K 4s electrons (see dashed lines in Fig. 1). At low
doping ($<$0.0066 e$^-$/C, C/K=150), the conduction band of the
metallic SWCNT within the SWCNT bundles is subsequently doped.
Interestingly, the conduction band of the semiconducting SWCNT
(S$_1^\star$ vHs) is not filled for these doping levels. In the
photoemission response this leads to a parallel shift of in the
position of the S$_1$ and S$_2$ peak up to 0.3 eV to a higher
binding energy. For the corresponding metallic SWCNT within the
bundle the shift of the $M_1$ peak is lower than for the
semiconducting SWCNT. At higher doping ($>$0.008 e$^-$/C, C/K=125)
the S$_1$ peak shifts upward beyond the original position of the
S$_2$ peak and the corresponding S$_1^\star$ vHs is occupied. As
can be seen in the figure for further increasing the doping level
the peaks of the vHs are smeared out and finally disappear
completely ($>$0.02 e$^-$/C, C/K=50). This can be explained by
effects like an increasing number of scattering centers (K$^+$
counter ions) and by an increasing inter-tube interaction within
the SWCNT bundle in the intercalation compound. The overall shape
of the spectra of this highly doped samples are also very similar
to the corresponding GIC \cite{oel}. The Fermi level shift can be
extracted from the shift of the $\pi$ band at 3 eV. For the sample
with C/K=15 we observe $\Delta E_F=1$ eV which is consistent with
the 1.25 eV shift observed for the GIC KC$_8$ \cite{oel} and in
good agreement with the above mentioned core level shifts.
Notably, none of the doping levels exhibit peaks corresponding to
the former unoccupied S$_1^\star$, S$_2^\star$, and M$_1^\star$
vHs. On the other hand, for all highly doped samples a satellite
in the photoemission response occurs at about 200 meV which is
very close to the frequency of the G-Line of the doped SWCNT
\cite{peaprl}. Hence, it is tempting to explain this as a
redistribution of the spectral weight by electron phonon coupling.
This explanation is also supported by the close analogy of the
line shape to the low temperature photoemission spectra of the
metallic C$_{60}$ intercalation compound K$_3$C$_{60}$ which is
dominated by strong satellites due to coupling to phonons and to
the charge carrier plasmon \cite{c60}. This change in the
line-shape due to electron-phonon coupling also explains the
absence of the above-mentioned photoemission peaks which are
related to the S$_1^\star$, S$_2^\star$, and M$_1^\star$ vHs.

\begin{figure}[htb]
\includegraphics*[width=0.6\linewidth]{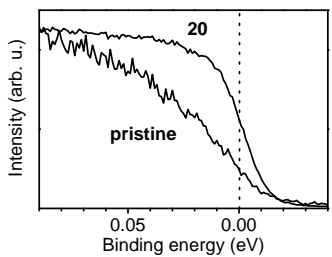}
\caption{Valence band photoemission spectra in the vicinity of the
Fermi level (dotted line) for pristine (upscaled for clarity) and
highly doped (C/K=20) SWCNT mats.}
\end{figure}

\begin{figure}[htb]
\includegraphics*[width=0.68\linewidth]{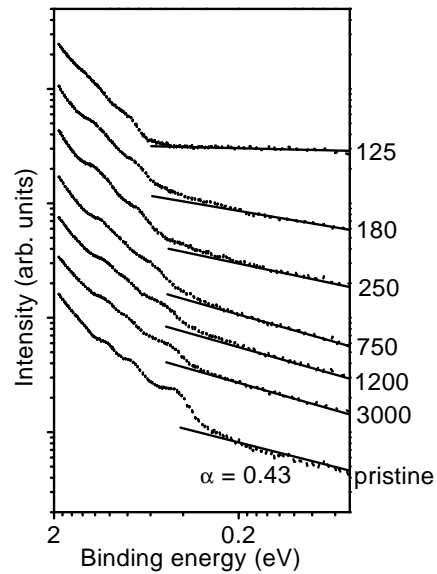}
\caption{Double-logarithmic representation of the photoemission
spectra for the analysis of the power law scaling within TLL
theory close to the chemical potential. $\alpha$ results from a
linear fit between 0.05 and 0.2 eV  (lines in the figure) and is
also shown for the pristine case.}
\end{figure}

We now turn to the analysis of the low energy electronic
properties (below 0.3 eV binding energy). For a three dimensional
system one would expect a constant DOS and a typical Fermi edge.
As can easily be seen in Fig. 2, this is only observed at very
high doping levels. For the pristine SWCNT sample, on the other
hand, there is a strong suppression in the DOS near the Fermi
energy. This behavior can be fully explained within a TLL theory
of one dimensional metals. A key manifestation of the TLL state is
the renormalization of the DOS (n(E)) near the Fermi edge which
shows a power law dependence n(E) $\propto$ E$^\alpha$ where
$\alpha$ depends on the size of the Coulomb interaction and can be
expressed in terms of the Luttinger parameter g as
$\alpha=(g-g^{-1}-2)/8$ \cite{egger,kane,egger2}. In the case of
photoemission, $\alpha$ can be directly derived from a linear fit
of the double-logarithmic representation of the response
 at low binding energy (see Fig. 3). For the pristine SWCNT (bottom curves in Figs. 1, 3) we
observe a power law scaling $\alpha=0.43$, g=0.18 which is -
within the experimental error - identical to the previously
reported value ($\alpha=0.46-0.48)$ \cite{tll_nature}.

\begin{figure}[htb]
\includegraphics*[width=0.68\linewidth]{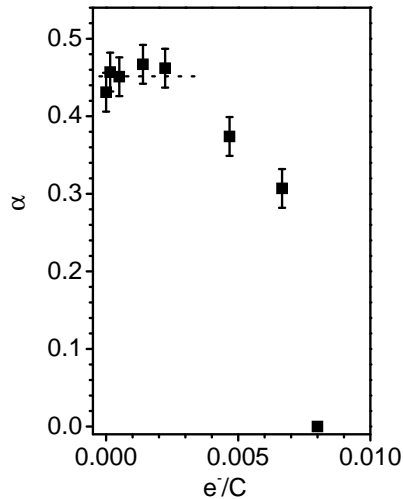}
\caption{Power law scaling factor $\alpha$ as a function of
different doping levels (e$^-$/carbon).}
\end{figure}

\par
These results give rise to the important question as to what
happens to the TLL as a function of doping and especially at what
doping level the transition from a TLL ground state to a FL
behavior occurs. Fig. 3 shows typical examples for the TLL scaling
for doping levels up to C/K=125. The lines in the figure represent
the linear fit for the determination of the scaling factor
$\alpha$. It can be easily seen that $\alpha$ depends on the
doping level and vanishes at C/K=125. The details in the doping
dependence of $\alpha$ are depicted in Fig. 4. For doping levels
$<$0.003 e$^-$/C no change in the TLL parameter is observed within
experimental error. As mentioned above, up to this doping level
the shift of the S$_1$ peak to a higher binding energy is 0.3 eV
which is small enough not to start any filling of the S$_1^\star$
level and only the metallic SWCNT within the bundle are doped.
Hence, the metallic SWCNT remain TLL upon filling of the
conduction band and a corresponding Fermi level shift up to 0.3
eV. This also means that the long range Coulomb interaction is
essentially unaffected by the potential of the counter ions. This
is in good agreement with predictions showing that for a TLL the
power law scaling parameter $\alpha$ is not affected until the
first vHs M$_1^\star$ is reached and additional conduction
channels are possible \cite{egger2}.
\par
For the intermediate doping levels ($0.003 < 0.008 $ e$^-$/C) one
starts to dope the semiconducting SWCNT. This can be substantiated
by two facts. Firstly, the additional shift of the S$_1$ peak over
this doping range is only 0.07 eV, namely from 0.3 to 0.37 eV.
Secondly, the photoemission spectral weight close to the Fermi
level strongly increases (see also Fig. 1). Regarding the TLL
scaling, $\alpha$ decreases to 0.35 (0.0046 e$^-$/C), then yet
further to 0.3 (0.0066 e$^-$/C) and finally shows a rapid
transition to zero at 0.008 e$^-$/C. At this doping level the
Fermi level is within the S$_1^\star$ vHs for the majority of the
semiconducting SWCNT. We can explain this observation as a
transition from a sample with roughly 1/3 of metallic SWCNT with
TLL behavior to a sample which consists of only metallic SWCNT and
has the scaling behavior of a normal FL. At even higher doping
levels a Fermi edge is observed. This can be explained by the fact
that at this dopant concentration the Coulomb interaction between
charge carriers is strongly screened by neighboring tubes, now
being all metallic, and/or by the fact that in a sample of only
metallic tubes the hopping matrix element between the tubes is
strongly enhanced, thus yielding a more three-dimensional
electronic structure \cite{ll_rev}.
\par
In summary, we have studied the character of the electron liquid
of bundles of SWCNT as a function of dopant concentration, i.e. as
a function of the position of the Fermi level. As long as the
Fermi level shift is small enough to only affect the states of the
metallic tubes, a TLL behavior is observed, indicating a weak
interaction between the individual metallic tubes within a bundle.
When the Fermi level is shifted into the states of the
semiconducting tubes, the reduced power law behavior indicates a
transition to a FL.  This can be explained by a more three
dimensional band structure and/or screening effects which lead to
a normal FL behavior of the entire bundle. Thus, we have expounded
for the first time a doping induced transition from a
quasi-one-dimensional system with a TLL behavior to a
intercalation compound with normal FL behavior.

\par
{\bf Acknwledgements:} This work was supported by the DFG PI 440.
We thank S. Leger, R. H\"ubel, and K. M\"uller for technical
assistance. H.K. acknowledges for a support by Industrial
Technology Research Grant Program in '03 from New Energy and
Industrial Technology Development Organization (NEDO) of Japan.

\end{document}